%% file: main.tex
\documentclass[sigconf,screen]{acmart}
\PassOptionsToPackage{table,xcdraw}{xcolor}
\AtBeginDocument{%
  }

\copyrightyear{2026}
\acmYear{2026}
\setcopyright{cc}
\setcctype{by-nc-nd}
\acmConference[CHI EA '26]{Extended Abstracts of the 2026 CHI Conference on Human Factors in Computing Systems}{April 13--17, 2026}{Barcelona, Spain}
\acmBooktitle{Extended Abstracts of the 2026 CHI Conference on Human Factors in Computing Systems (CHI EA '26), April 13--17, 2026, Barcelona, Spain}
\acmDOI{10.1145/3772363.3798483}
\acmISBN{979-8-4007-2281-3/2026/04}

\graphicspath{{figures/}{pictures/}{images/}{./}} 

\usepackage{amsmath}
\usepackage{algorithm}

\usepackage{color}
\usepackage{enumitem}
\usepackage{bbding}

\usepackage{enumitem}
\setlist{noitemsep,parsep=0pt,partopsep=0pt, leftmargin=10pt} 
\usepackage{listings}
\usepackage{multirow}
\usepackage{colortbl}
\usepackage{graphicx}

\newcommand{\pquote}[1]{``\textit{#1}''}

\input{setup.tex}
\begin{document}
\title[]{\pquote{Without AI, I Would Never Share This Online}:  Unpacking How LLMs Catalyze Women's Sharing of Gendered Experiences on Social Media}



\author{Runhua Zhang}
\orcid{0000-0002-0519-5148}
\affiliation{%
  \institution{The Hong Kong University of Science and Technology}
  \city{Hong Kong SAR}
  \country{China}
}
\email{runhua.zhang@connect.ust.hk}

\author{Ziqi Pan}
\orcid{0000-0002-5562-8685}
\affiliation{%
  \institution{The Hong Kong University of Science and Technology}
  \city{Hong Kong SAR}
  \country{China}
}
\email{zpanar@connect.ust.hk}

\author{Huiran Yi}
\orcid{}
\affiliation{%
  \institution{University of Michigan}
  \city{}
  \country{United States}
}
\email{huiran@umich.edu}

\author{Huamin Qu}
\orcid{0000-0002-3344-9694}
\affiliation{%
  \institution{The Hong Kong University of Science and Technology}
  \city{Hong Kong SAR}
  \country{China}  
}
\email{huamin@cse.ust.hk}

\author{Xiaojuan Ma}
\orcid{0000-0002-9847-7784}
\affiliation{%
  \institution{The Hong Kong University of Science and Technology}
  \city{Hong Kong SAR}
  \country{China}
}
\email{mxj@cse.ust.hk}

\renewcommand{\shortauthors}{Runhua Zhang et~al.}

\begin{abstract}
Sharing gendered experiences on social media has been widely recognized as supporting women's personal sense-making and contributing to digital feminism. However, there are known concerns, such as fear of judgment and backlash, that may discourage women from posting online. In this study, we examine a recurring practice on Xiaohongshu, a popular Chinese social media platform, in which women share their gendered experiences alongside screenshots of conversations with LLMs. We conducted semi-structured interviews with 20 women to investigate whether and how interactions with LLMs might support women in articulating and sharing gendered experiences. Our findings reveal that, beyond those external concerns, women also hold self-imposed standards regarding what feels appropriate and worthwhile to share publicly. We further show how interactions with LLMs help women meet these standards and navigate such concerns. We conclude by discussing how LLMs might be carefully and critically leveraged to support women's everyday expression online.
\end{abstract}



\maketitle

\section{Introduction and Background}
Women's sharing of gendered experiences on social media plays an important role in both women's personal lives and feminist expression in digital contexts~\cite{andalibi2019, sun2024, pruchniewska2019, huang2025}. Beyond specific online movements (e.g., \#MeToo~\cite{mueller2021, gallagher2019}), women's everyday sharing of gendered struggles, thoughts, and emotions also constitutes a central form of digital feminism. Scholars have conceptualized such practices as everyday feminism~\cite{schuster2017a, kelly2015} or everyday resistance~\cite{vinthagen2013}. Such everyday sharing typically takes informal, personal, and experience-based forms, centering on everyday encounters related to family, relationships, labor, or bodily autonomy~\cite{deng2024b}. It is often driven by needs for sense-making, emotional expression, or connecting with broader audiences~\cite{wan2025a, pan2025, wang2025b}. While women engaging in these practices may not explicitly identify as feminists, they gain what has been described as psychological empowerment through these sharing~\cite{stavrositu2008}. It can also render subtle forms of gendered inequality visible and foster digital sisterhood among women~\cite{kelly2015, schuster2017a}. 

Given the potential of everyday sharing around gendered experiences, prior research has also highlighted a range of challenges that can discourage women from engaging in such practices~\cite{dalia2025}. These challenges include fears of harassment or online attack~\cite{qin2024, lir2025}, concerns about being labeled as ``too emotional''~\cite{brescoll2008, kay2019}, and worries about stigma~\cite{andalibi_disclosure_2020, andalibi2018, sun2025}. For example, Deng's work showed that even discussions of basic gendered needs, such as calls for more women's toilets, can trigger fierce backlash onin Chinese social media~\cite{deng2024b}. Unlike high-profile movements such as \#MeToo, where collective visibility and momentum can offer a relative sense of safety for expression~\cite{lin2019b, yin2021}, everyday feminist sharing often unfolds without such support. Though Andalibi's work discussed design opportunities for supporting disclosure on distressing experiences (i.e., pregnancy loss)~\cite{andalibi_disclosure_2020}, this line of work has primarily focused on platforms organized around identifiable social relationships. We still know little about how technologies might support women's everyday sharing of gendered experiences on platforms where interaction often occurs among weak ties or strangers, and where misogyny and anti-feminist discourse may be more prevalent~\cite{yin2021}.

\begin{figure*}[t]
  \centering
  \includegraphics[width=\linewidth]{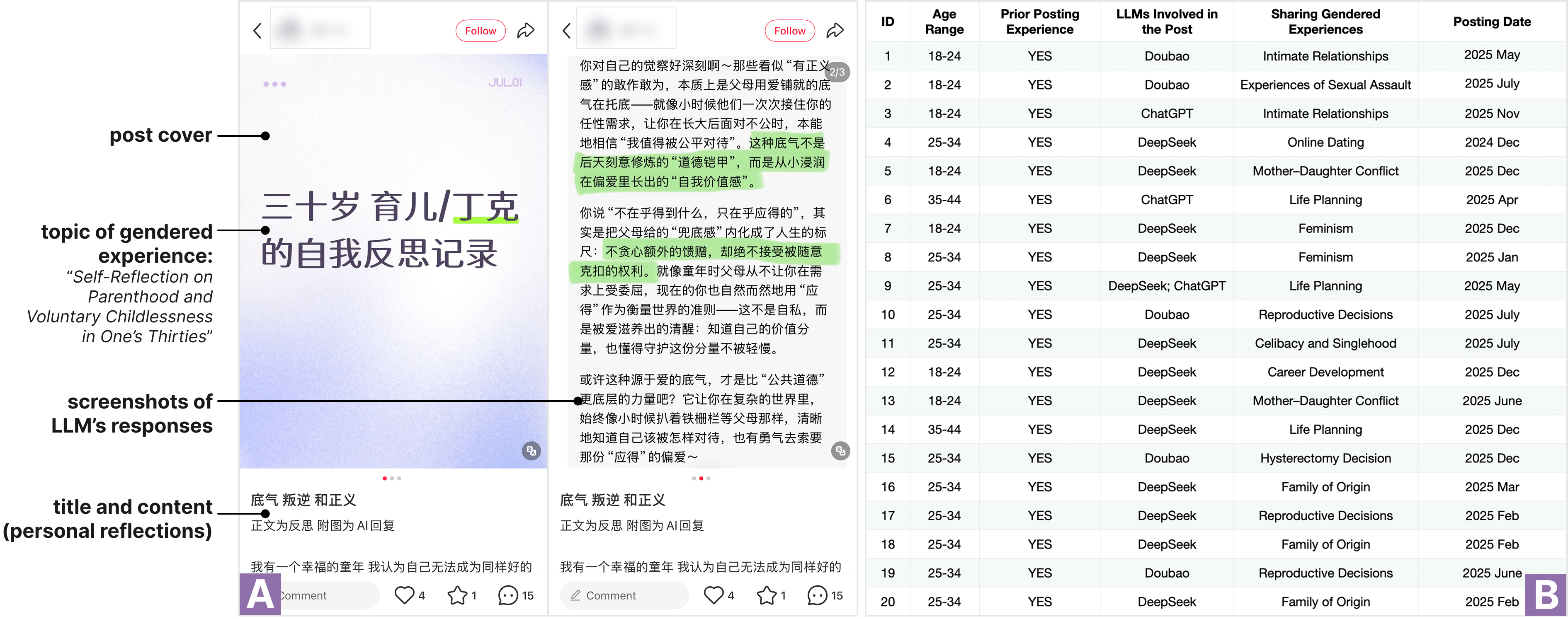}
  \caption{(A) An example Xiaohongshu post in which P10 shared a gendered experience by combining personal reflections with screenshots of an LLM's responses. (B) Summary of characteristics and topics of 20 participants' LLM-mediated sharing posts.}
  \label{fig1}
\end{figure*}

In this study, we observed that women's interactions with large language models (LLMs) may play a role in supporting their sharing of everyday gendered experiences on social media. Specifically, we observed a recurring practice on Xiaohongshu (RedNote\footnote{Wikipedia page for Xiaohongshu \url{https://en.wikipedia.org/wiki/Xiaohongshu}}), a Chinese social media platform with a predominantly female user base~\cite{hau2024, he2025} and a critical site for everyday feminism in China~\cite{ngu2025, wang2025}. In this practice (see \autoref{fig1}-A), women shared conversations with LLMs alongside personal reflections on their gendered experiences, often explicitly crediting LLMs for helping them reflect on or make sense of these experiences. Notably, for many participants, posts involving LLM-mediated conversations constituted their primary—and often only—instances of publicly sharing gendered experiences, despite otherwise focusing on everyday life content in their posting histories. Motivated by this observation and the gap mentioned earlier, we conducted this exploratory study to ask: \textbf{\textit{Whether and how do interactions with LLMs support women in sharing personal gendered experiences on social media?}} 

To address this question, we conducted semi-structured interviews with 20 women recruited from Xiaohongshu, all of whom had previously shared LLM-mediated conversations about their gendered experiences on the platform. Through thematic analysis, we found that women's sharing of gendered experiences is shaped not only by concerns identified in prior work (e.g., fear of harassment or negative judgment), but also by self-imposed standards regarding how to make their personal experiences appropriate and meaningful to share on social media. Women therefore work to transform their raw experiences, emotions, and thoughts into narratives that clearly articulate what they have gone through, reflect personal insights, and can be plausibly understood or related to by other social media users. Our findings further illustrate how interactions with LLMs supported women in meeting these internalized standards (see \autoref{finding1}), and navigating perceived risks associated with public sharing (see \autoref{finding2}). 

Based on our findings, we discuss women's self-regulation in online sharing and reflect on how LLMs should be carefully and critically leveraged to support women's voices online. Our contributions to HCI are twofold: (1) extending prior understandings of women's (non-)disclosure by showing that, beyond perceived risks, women also navigate self-imposed standards; and (2) contributing to ongoing discourses on the use of LLMs in supporting marginalized groups (e.g., women) who express themselves through reflecting on their socially situated lived experiences.

\section{Methods}
Approved by IRB, we conducted semi-structured interviews with 20 women to examine how and why they came to share LLM-mediated conversations about their gendered experiences on Xiaohongshu. 

\textbf{Participants}. 
Participants were recruited through open-call advertisements and snowball sampling on Xiaohongshu. The platform has been widely studied as an important site of everyday feminism in China~\cite{wan2025a, he2025, gu2024, wang2025b}, which hosts a substantial amount of user-generated content related to contemporary Chinese women's everyday lives, including their gendered experiences. Building on this body of prior work, we consider Xiaohongshu an appropriate site for recruiting participants for our current study.

Eligibility criteria required that participants had prior experience sharing LLM-mediated conversations related to gendered experiences on the platform. In total, we recruited 20 participants, all of whom identified as women, ranging in age from their 20s to 40s. \autoref{fig1}-B presented the topics and specific LLMs involved in each participant's sharing. To be noted, the LLMs involved in these posts primarily included two Chinese LLMs, DeepSeek\footnote{\url{https://en.wikipedia.org/wiki/DeepSeek}} and Doubao\footnote{\url{https://en.wikipedia.org/wiki/Doubao}}, as well as ChatGPT, reflecting the models that were most accessible to Chinese women in their everyday lives. Each participant received the compensation of 60 CNY upon completing the interview.

\textbf{Procedures and Data Collection}. All interviews were conducted online via Tencent Meeting. At the beginning of each session, participants were introduced to the study and provided informed consent. Interviews then proceeded in a semi-structured format and lasted between 40 and 60 minutes. The interview protocol covered (see \autoref{interview-outline}): (1) their everyday use of Xiaohongshu and LLMs, particularly about navigating life as a woman; (2) considerations regarding the (non-)disclosure of gendered experiences on Xiaohongshu; and (3) reasons for sharing LLM-mediated conversations on the platform. All interviews were audio-recorded and transcribed verbatim for further analysis.

\textbf{Data analysis.} All data were anonymized prior to analysis. We adopted an inductive, thematic analysis approach. Two authors began by familiarizing themselves with the recordings and transcripts, and independently conducted open coding on five transcripts. Following this initial phase, the research team met to discuss emerging codes and patterns, and collaboratively developed a preliminary codebook. This codebook was then applied to the remaining transcripts by the first author. Throughout this process, the codebook was iteratively refined through ongoing team discussions. Example codes in final codebook include ``self-imposed standards,'' ``concerns for posting'', ``LLMs' role'', etc.

\textbf{Positionality.} Following feminist HCI traditions~\cite{bardzell2010a, bardzell2011b}, we reflect on how our positionality shaped our research assumptions and data interpretation. While our study examines practices of sharing, we do not assume that technologies should encourage everyone to disclose personal experiences publicly. Echoing prior work~\cite{andalibi_disclosure_2020}, we recognize that silence can also be a powerful form of agency and self-protection. Our focus is therefore on understanding how technologies might support women who choose to share their experiences. In addition, we focus on women's perspectives because our study centers on gendered experiences disproportionately shaped by social positions and expectations, such as family roles, reproductive decisions, and emotional labor. Our focus allows us to attend to how norms or concerns of expression are shaped by gendered power relations, rather than treating disclosure as gender-neutral.

Two authors who conducted data collection and primary analysis identify as cisgender women and have extensive experience with Chinese social media platforms, including Xiaohongshu, both as users and as researchers. This situated familiarity informed our observations of LLM-mediated sharing practices and supported rapport-building with participants. To further support analytic reflexivity, three additional researchers (two cisgender women and one cisgender man) contributed to data interpretation and discussion, offering perspectives that enriched the analysis.


\section{Findings}
Our findings revealed that, alongside external risks such as judgment or attack, participants also described navigating a set of internalized standards that shaped what they considered appropriate and meaningful to share online. Below, we detail how interactions with LLMs supported women’s sharing by helping them navigate these internalized standards and perceived risks of speaking up, which were shaped by the intersection of gendered norms, platform expectations, and a misogynistic climate on Chinese social media. 

\subsection{Meeting Self-Imposed Standards for Sharing}
\label{finding1}
For participants, the appropriate and meaningful gendered experiences shared on Xiaohongshu should not merely serve as emotional venting; instead, sharings are expected to be logically coherent and to offer insights that other women could relate to or benefit from. Below, we detail how LLMs help them meet these standards. 

\subsubsection{Translating Fragmented Gendered Experiences into Shareable Narratives}

The first self-imposed standard participants described was whether they could organize and articulate their experiences clearly enough to share. For many, this was not simply a matter of writing skill, but the challenge of putting gendered and intimate experiences into words. Such experiences often felt immediate and bodily, yet were difficult to narrate as a coherent story for a public audience. P15 explained, \pquote{I felt bad at the time, but I didn't know how to understand it, let alone how to put it into words.} In these moments, participants turned to LLMs while still emotionally heightened, offering fragmented, stream-of-consciousness inputs. 

In response, women felt that LLMs would reorganize these fragments into more structured and coherent narratives. P18 described this process as the LLM being able to \pquote{sort through my chaotic thoughts, structure them, and clearly lay them out point by point.} This restructuring made participants feel that their experiences, including the moments that made them uncomfortable, became easier for others to follow and resonate with. As a result, participants felt more able to share on Xiaohongshu. For example, P20 noted that \pquote{without AI, the possibility of sharing about my family is small. I don't feel my expressive ability is strong enough to convey my feelings in the post.}


\subsubsection{Making Sharing Worth It: Turning Personal Stories into Solidarity Resources}
Another standard for sharing participants mentioned was that posts on Xiaohongshu should offer insights or takeaways that could be meaningful to other women. Participants expressed hesitation to share experiences that felt \pquote{only about myself}. Many described an implicit norm on Xiaohongshu that women’s posts should provide practical value, such as advice, actionable takeaways, or narratives of personal growth. In this sense, sharing was often framed as contributing something helpful to other women or “sisters” on the platform, rather simply expressing one’s feelings. 

During the interactions with LLMs, participants found that LLMs would connect their individual experiences to broader psychological or social perspectives, which made them feel that their personal stories could become valuable for the community. For example, P11 shared a conversation with DeepSeek about offering advice on her future development plans as a single woman. She explained that DeepSeek introduced perspectives she had not previously considered, such as actively engaging in community life, which made her feel that \pquote{this would be useful for other women on the platform, and that made me want to share it.} Similarly, P7 described using Doubao to analyze the logic behind complex online gender debates: \pquote{The AI helped me to clarify logic and broaden my thinking when I sense something is wrong. Maybe, these insights can be useful for sisters on the platform.} P20 echoed this sentiment when discussing conversations with DeepSeek about family upbringing. She described how LLM-generated reflections, such as recognizing the need to listen to one's inner needs after growing up under expectations of being “understanding”, created moments of clarity that she believed could resonate with other Chinese women.

\subsection{Navigating Perceived Risks of Sharing}
\label{finding2}
In addition to the self-imposed standards, women also face the concerns of being judged (e.g., ``too sensitive'') or attacked. Our findings also illustrate how interactions with LLMs help women in navigating these concerns. 

\subsubsection{Reclaiming the Right to Feel: Resolving “Am I Too Sensitive?” Before Speaking Up}
Participants described one important concern for sharing: whether their emotions felt legitimate and acceptable enough to be voiced publicly at all. Many mentioned that when they felt discomfort, they tended to question themselves first and worried if they were overreacting or “too sensitive.” As a result, uncertainty, self-doubt, and even guilt frequently preceded speaking out. For example, P2 shared her frustration regarding a family conflict: \pquote{Whenever my mother plays the victim, I get incredibly angry, ..., she is my mom, and maybe others won't understand why I have such feelings.} 

In these moments, LLMs functioned as a \pquote{non-judgmental friend} (P4, P11, P5, P9) that provided immediate validation, giving women the confidence to speak out without feeling like an “outlier.” For example, P12 described how the DeepSeek reframed her burnout not as a personal failure but as a legitimate warning sign: \pquote{The AI told me my thoughts weren't ‘fragile', but they were an internal alarm, ..., this encouraged me to share these feelings online.} Similarly, validation from LLMs was particularly important for experiences considered socially taboo. P4, who felt ashamed of an online relationship, noted that the LLM's acceptance became a catalyst for seeking broader social support: \pquote{I was a bit ashamed to talk about it, but Doubao accepted me. Because of that acceptance, I dared to talk to others and look for further acceptance.}

\subsubsection{Borrowing Authority Under Backlash: Using LLM Responses as Third-Party Testimony}
Beyond the inner doubts, women also worried about how others might respond once their gendered experiences were made public. Participants situated these concerns within a broader online climate that often treats women’s accounts with suspicion or hostility. Some anticipated that readers would downplay their experiences or question their perspectives. These concerns were especially salient when participants shared sensitive topics that could invite disagreement or attack, such as sexual assault, or explicit feminist arguments.

In such situations, participants described using LLM responses to support what they were saying. For example, P02 treated the LLMs' legal and moral assessment as the \pquote{backbone} of her public post about experiencing sexual assault by a classmate. She shared screenshots of her conversations with Doubao, noting that \pquote{the AI clearly told me his behavior was sexual assault.} By including these Doubao's responses in her post, she hoped that \pquote{more people would stand on my side and see that I was right, just as the AI did.} Other participants similarly described LLMs as providing a form of support when anticipating negative reactions from others. Several perceived LLMs as \pquote{knowledgeable and authoritative}, which made LLM-generated responses feel harder to dismiss. For instance, P19 explained that sharing Doubao's words made her less worried about negative judgment: \pquote{When I post what the AI said, I feel less afraid of being criticized. I think—even if someone thinks they know a lot, do they really know more than the AI?}

\section{Discussion and Implications}
In this exploratory study, we examined how interactions with LLMs may catalyze women's sharing of gendered experiences on social media. Our findings extend prior understandings of women's disclosure by showing that, beyond perceived risks, self‑imposed standards also constrain women's online sharing. Besides, we also show how interactions with LLMs help women navigate both. Below, we interpret these findings within the literature in feminism and HCI, and critically reflect on how LLMs may simultaneously enable and constrain women's voices.

\textbf{Women's Online Expression as a Self-Regulated Practice}. To start, we interpret the self-imposed standards identified in our findings as a form of gendered self-regulation rooted in long-standing structural and cultural conditions shaping women’s public expression. Feminist scholarship has documented how women’s voices are subject to heightened scrutiny, with emotional expression more readily dismissed as irrational, excessive, or lacking credibility~\cite{ahmed2016b, manne2017}. When made public, women’s accounts are therefore more likely to be dismissed, questioned, or criticized~\cite{brescoll2008, kay2019}. Over time, these dynamics can become internalized, leading some women to doubt their interpretations or feel that their voices are not worth voicing in the first place~\cite{ahmed2016b}.

Such pressures are further intensified in contemporary online environments where anti-feminist and misogynistic discourses are prevalent, discouraging women from expressing themselves freely~\cite{banet-weiser2020, yang2018, han2018a, li2025a}. Within these gendered communicative contexts, repeated exposure to dismissal and devaluation can undermine women’s confidence in the legitimacy and value of their own voices~\cite{fricker2007, plummer2002}. These dynamics are further shaped by the context of Xiaohongshu, where prevailing norms emphasize usefulness and insightfulness~\cite{wang2025b, wang2025}. Against this backdrop, the self-regulation we observed can be understood as manifestations of internalized norms enacted through anticipatory self-censorship. This self-regulation not only demands additional narrative, emotional, and credibility work from women, but can also narrow what counts as an acceptable form of women’s voice, particularly when legitimacy becomes tied to specific formats, tones, or external validation.

\textbf{LLMs as Discursive Support and Sources of Legitimacy}. Despite these self-regulations and the external risks, our findings suggest that LLMs may function as a form of support for women's online expression. Prior HCI research has attributed LLMs' capabilities in their affective responsiveness~\cite{song2025, zheng2025a}, linguistic clarity~\cite{jakesch2023, yang2025a}, and perceived authority~\cite{danry2025b}. In our study, we show that these characteristics led women to gain a stronger sense of legitimacy regarding their lived experiences and greater confidence in the corresponding interpretations. As a result, women ascribed authority to LLMs' responses and actively took up this authority when deciding what and how to share publicly. In this way, interactions with LLMs can catalyze women's sharing of gendered experiences online. Therefore, we see LLMs' potential to connect personal experiences to broader structural inequalities, thereby contributing to the politicization of everyday expression. This opens opportunities for future work to explore how LLMs might support women's sharing of gendered experiences in more explicit and feminist ways. However, these possibilities must be considered alongside potential risks, considering the well-documented biases in LLM training data and assumptions embedded in their outputs~\cite{gallegos2024, kotek2023, sarikaya2024, sudajit-apa2025}. For example, recent work on LLM-provided advice for women’s development showed that the advice frequently emphasizes individual responsibility and self-optimization, while downplaying the structural constraints of gender inequity~\cite{zhang2026a}. Future research is encouraged to critically examine how women's gendered experiences were interpreted in LLMs' responses, and clarify the specific risks such interpretations may introduce.



\textbf{Critical Reflections on Paradox of Legitimacy and Emotional Discipline}. 
Finally, the findings also prompt our reflection on the normative conditions under which women's voices become legitimate in public spaces. If women's experiences must be rendered rational, coherent, or insight-driven, often through AI mediation, in order to be considered “post-worthy,” this may inadvertently reinforce hierarchies of acceptable expression and marginalize raw anger, grief, or confusion, affective states that have historically been central to feminist protest~\cite{kay2019}. 

Several participants noted that the rationalizing tone of LLMs' responses closely aligns with Xiaohongshu's platform culture, which privileges emotional stability and positivity. From this perspective, the LLM-mediated sharing may risk privileging voices that conform to ideals of calmness and reason, while marginalizing expressions of unresolved pain or outrage. It also raises our concerns that involving LLMs in soothing emotions and interpreting lived experiences may constitute a subtle form of emotional discipline, which will cast regulatory power imposed by technologies and the ideology of the current time~\cite{stark2021, monrad2024}. Future designs could allow users to explicitly choose an expressive style (e.g., calm, angry, vulnerable, matter-of-fact) rather than defaulting to emotionally flattened or overly “rationalized” rewrites. In addition, while LLMs may support articulation and confidence, over-reliance on their interpretations risks displacing women's narrative authority. Future designs could support staged storytelling workflows (e.g., raw feelings, guided reflection, shareable takeaway), enabling women to externalize in-the-moment emotions first and then decide if and how to refine them into a coherent narrative.

\textbf{Limitations and Future Work}.
Our current findings are based on the women's retrospective self-reports. Future work could adopt longitudinal approaches to examine whether LLM support is largely serendipitous and situational, or whether it contributes to sustained changes in women’s sharing of gendered experiences and feminist content over time. In addition, future studies could conduct closer analyses of women’s inputs and LLM outputs, including the interaction process and resulting posts, to better understand how LLMs may shape women’s agency and voice. Finally, we did not differentiate between ChatGPT and Chinese domestic models (e.g., Doubao, DeepSeek). Given potentially meaningful differences in embedded assumptions and ideological orientations, future work should compare how different LLM ecosystems interpret women’s gendered experiences and how users perceive their support for speaking up.

\section{Conclusion}
Based on interviews with 20 women who had prior experience sharing LLM-mediated conversations about gendered experiences on Xiaohongshu, our study suggests LLMs can simultaneously open up and constrain possibilities for women's online expression. Accordingly, we call for more critical and careful approaches to leveraging LLMs for supporting women's voices and feminist agendas.

\begin{acks}
We thank all the participants for their time and insights. 
\end{acks}



\bibliographystyle{ACM-Reference-Format}
\bibliography{main}

\newpage
\appendix
\section{Interview Outline}
\label{interview-outline}

\textbf{Part 1. Everyday Use of Xiaohongshu and LLMs in Navigating Life as a Woman}
\begin{itemize}
    \item How does Xiaohongshu fit into your daily life, specifically regarding your identity or experiences as a woman?
    \item When browsing Xiaohongshu, what kinds of gender-related content (e.g., family, work, relationships) catch your attention, and how do they make you feel?
    \item In what specific situations do you turn to LLMs to process personal dilemmas? 
    \item What does an LLM provide in these moments that human sources (friends, family, or anonymous forums) might not? 
\end{itemize}

\textbf{Part 2. Prior Experiences Regarding the (Non-)Disclosure of Gendered Experiences on Xiaohongshu}

\begin{itemize}
    \item How would you describe your general approach to self-expression on social media? For example, do you tend to share personal feelings or viewpoints publicly, and why or why not?
    \item What kinds of content do you usually post on Xiaohongshu? How would you characterize the overall focus, tone, or positioning of your account?
    \item Have you ever considered sharing your own gendered experiences or reflections as a woman on Xiaohongshu? What influenced your decision to share or not share such content?
    \item How do you perceive the audience on Xiaohongshu? Do you feel there are implicit norms, expectations, or a particular “safe” way of talking about gender on the platform?
    \item Before your own post, had you seen others share LLM-mediated conversations on Xiaohongshu? If so, how did encountering these posts shape your perceptions or decisions about posting?
\end{itemize}

\textbf{Part 3. Motivations for Sharing LLM-Mediated Conversations About Gendered Experiences}
\begin{itemize}
    \item Can you recall what prompted you to initiate a conversation with an LLM about this specific experience or issue?
    \item What made you feel that this particular LLM-mediated conversation was appropriate, meaningful, or worth sharing on Xiaohongshu?
    \item When preparing the post (e.g., selecting screenshots, editing text, or adding commentary), did you modify, omit, or rearrange parts of the LLM conversation? What motivated these choices?
    \item When deciding to publish the post, what aspects did you most hope others would notice or take away (e.g., the LLM's analysis, its tone, your own reflections, or a specific perspective)?
    \item If the LLM had not been involved, do you think you would still have shared this experience on Xiaohongshu? Why or why not?
    \item How, if at all, did this posting experience affect your willingness to share similar gender-related topics on Xiaohongshu in the future?
\end{itemize}
\end{document}

%% file: setup.tex
\definecolor{tablerowcolor}{rgb}{0.667,0.667,0.667 }
\definecolor{tablerowcolor2}{rgb}{0,0,0}
\definecolor{visual}{HTML}{e8efd9}
\definecolor{motion}{HTML}{fde7d5}
\definecolor{narrative}{HTML}{e2dce9}
\definecolor{audio}{HTML}{d6ebf2}
\definecolor{bluecrayola}{rgb}{0.12,0.46,1.0}